\documentclass[twocolumn]{aastex701}

\usepackage{tabularx}
\usepackage{longtable}
\usepackage{lineno}
\usepackage{amssymb}
\usepackage{gensymb}
\usepackage{xcolor}

\begin{document}

\linenumbers

\title{A Cepheid-based Distance to the Type 1 Seyfert NGC\,3783}

\author[orcid=0009-0000-9670-2194]{Madison Markham}
\affiliation{Department of Physics and Astronomy, Georgia State University, Atlanta, GA 30303, USA}
\email{mmarkham@gsu.edu}

\author[orcid=0000-0002-2816-5398]{Misty C. Bentz}
\affiliation{Department of Physics and Astronomy, Georgia State University, Atlanta, GA 30303, USA}
\email{bentz@gsu.edu}

\author[orcid=0000-0002-8224-1128]{Laura Ferrarese}
\affiliation{NRC-Herzberg Astronomy and Astrophysics, 5071 West Saanich Road, Victoria, BC, V9E 2E7, Canada}
\email{Laura.Ferrarese@nrc-cnrc.gc.ca}

\author[orcid=0000-0003-0017-349X]{Christopher A. Onken}
\affiliation{Research School of Astronomy and Astrophysics, Australian National University, Canberra, ACT 2611, Australia}
\email{christopher.onken@anu.edu.au}

\author[orcid=0000-0001-9191-9837]{Marianne Vestergaard}
\affiliation{DARK, Niels Bohr Institute, University of Copenhagen, Jagtvej 155, DK-2200 Copenhagen}
\affiliation{Steward Observatory and Department of Astronomy, University of Arizona, 933 N Cherry Avenue, Tucson AZ 85721, USA}
\email{mvester@nbi.ku.dk}

\begin{abstract}

We present a Cepheid-based distance to the Type 1 Seyfert NGC\,3783 using the Hubble Space Telescope. Images were obtained over 15 visits using nonredundant time spacing, with ten observations in the F350LP filter and eight observations each in the F555W and F814W filters. We found 1541 unique point sources across all observations, and we fit Cepheid templates to analyze these light curves. From these data, we determined a full sample of seven Cepheid candidates. Of these seven candidates, four are the most convincing and comprise our ``strong" sample of Cepheid candidates. We fit the period$-$luminosity relationship to the total sample and to our ``strong" sample using the LMC calibration by \cite{riess_large_2019}. Adopting an independent distance constraint to the LMC given by \cite{pietrzynski_distance_2019}, and correcting for the potential effects of metallicity, we determine a distance modulus of $\mu = 33.118 \pm 0.154$ mag using our strong sample. This value is formally consistent with the value derived using our full sample and corresponds to a distance of $D = 42.03 \pm 2.98$ Mpc. NGC\,3783 is therefore among the farthest galaxies with a Cepheid-based distance measurement.

\end{abstract}


\section{Introduction} \label{sec:intro}

Henrietta Leavitt's work with Cepheid variable stars revolutionized the field of astronomy. By observing Cepheids in the LMC and SMC, she discovered that longer pulsation periods corresponded to brighter luminosities---thus giving rise to the Leavitt Law, or the period$-$luminosity relationship \citep{leavitt_periods_1912}. Though it was not appreciated at the time, this established Cepheids as standard candles in astronomy. It was over a decade later when Edwin Hubble applied Leavitt's work to Cepheids in the Andromeda galaxy \citep{hubble_cepheids_1925}, and the strength of this method was realized. In measuring the distance to the Andromeda galaxy, the scale of the Universe was unlocked.

Decades later, the success of the Hubble Space Telescope (HST) once again revolutionized extragalactic astronomy. The launch of HST was in part motivated by extragalactic Cepheids, which were used in the HST Key Project to determine the value of the Hubble constant \citep{kennicutt_measuring_1995,freedman_final_2001}. Since its launch, HST has been used to detect and characterize Cepheids in nearby galaxies with great success (e.g., \citealt{saha_cepheid_2006,riess_large_2019,yuan_cepheid_2020,bentz_cepheid-based_2019}). What was not planned, however, was the unexpected discovery that practically all massive galaxies host supermassive black holes (SMBHs) at their gravitational centers, and this opened a new field of research. The masses of these SMBHs, despite having gravitational influence over a small distance, have been found to correlate with properties of their host galaxies, including bulge velocity dispersion ($M_{\rm BH} - \sigma_{\star}$; \citealt{ferrarese_fundamental_2000, gebhardt_relationship_2000, gultekin_m-sigma_2009}) and bulge luminosity ($M_{\rm BH} - L_{\rm bulge}$; \citealt{kormendy_inward_1995, kormendy_coevolution_2013}). These relationships imply that AGN influence their host galaxies on massive spatial scales, and understanding the SMBHs within these systems is a key factor in understanding galactic evolution.

Reverberation mapping (RM) is a distance-independent technique used to measure the masses of SMBHs (e.g., \citealt{blandford_reverberation_1982,peterson_reverberation_1993}). The method trades spatial resolution for temporal resolution, measuring the time delay (or light-travel time) between various regions in an AGN and their response to variations in the central ionizing luminosity. When applied to the broad-line region (BLR), the time delay of an emission line coupled with the Doppler-broadened width of the line directly constrains the central black hole mass. By allowing time resolution to replace spatial resolution, RM has proven to be an incredibly useful technique in determining black hole masses of Seyfert 1 AGN. 

\begin{figure}[htb!]
    \centering
    \includegraphics[width=\columnwidth]{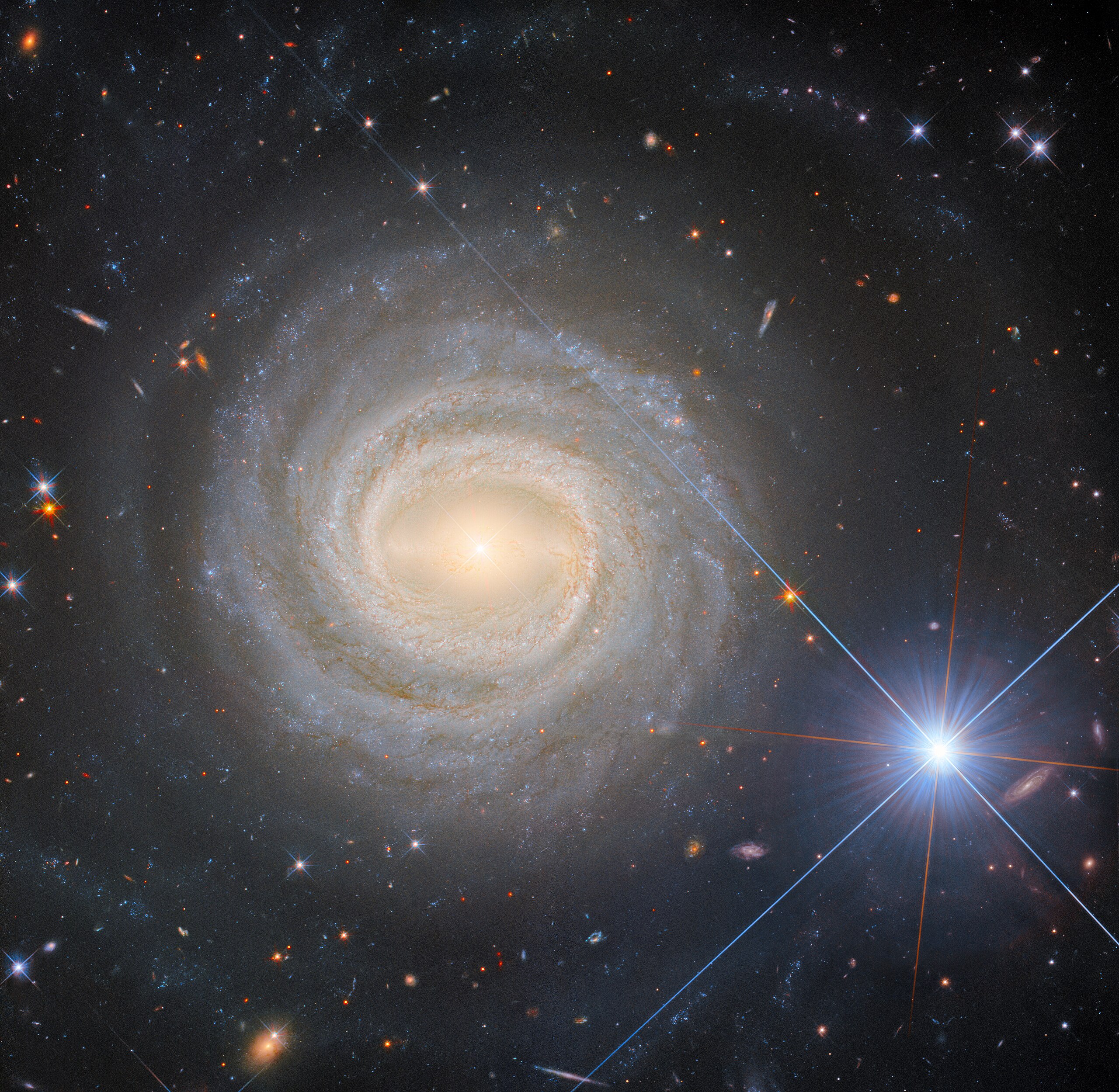}
    \caption{Multi-wavelength (F547M, F555W, F814W, F160W, F350LP) HST image of NGC\,3783. The field of view is $2.69' \times 2.63'$, and North is oriented $107.7 \degree$ counterclockwise from up. Credit: ESA/Hubble \& NASA, M.C.\ Bentz, D.J.V.\ Rosario.}
    \label{fig:pretty-pic}
\end{figure}

NGC\,3783 is a nearly face-on SBa galaxy with a strong bar located at $z = 0.00973$ (\citealt{de_vaucouleurs_third_1991,theureau_kinematics_1998}; see Figure \ref{fig:pretty-pic}). As one of the nearest and brightest Type 1 Seyferts ($\rm log \ \it{L_{\rm AGN}} \rm \approx 44.5 \ erg \ s^{-1}$, assuming $D = 38.5$ Mpc; \citealt{davies_insights_2015}), it is among the most well-studied AGN. NGC\,3783 has been a target for several RM studies, including simultaneous monitoring in the UV \citep{reichert_steps_1994} and optical \citep{stirpe_steps_1994}. \cite{bentz_robotic_2021} recently carried out a new RM campaign and was able to constrain velocity-resolved emission line responses. Dynamical modeling of the full reverberation response in NGC 3783 provided a direct and primary measurement of the black hole mass of $M_{\rm BH} = 2.82^{+1.55}_{-0.63} \times 10^7 \ M_{\odot}$ \citep{bentz_detailed_2021}. As a complement to RM measurements of the physical size of the BLR, NGC\,3783 is one of a select few nearby Type 1 AGN with an angular size of the BLR determined using interferometry \citep{gravity_collaboration_central_2021}. An accurate distance is required to convert the angular size of the BLR to a physical size for comparison with RM measurements.

Nearby AGN are important observational resources for resolved studies of physical processes, such as feeding and feedback, and these also rely on accurate distances. NGC\,3783 has been studied in detail across the wavelength spectrum. In the radio wavelength regime, \cite{shablovinskaia_alma_2025} found polarized synchrotron emission in NGC\,3783 that is likely due to shocks from outflowing material, and the physical extent of these shocks depends directly on distance. Dust surrounding the central AGN of NGC\,3783 has been observed in the mid-IR (e.g., \citealt{beckert_probing_2008,honig_dust_2013}), and these studies provide a better understanding of the structure of the torus. An accurate model of the torus is essential in modeling IR emission from an AGN, and therefore is necessary for modeling the spectral energy distribution (SED). Additionally, observations in the X-ray have revealed the presence of highly ionized outflows in NGC\,3783 (e.g., \citealt{mao_photoionized_2019,mehdipour_chasing_2017,mehdipour_delving_2025}), and probing the structure of these outflows requires a model of the central ionizing source. SEDs are constructed and used to model ionizing radiation from an AGN, and they are often used to determine AGN luminosity. However, AGN luminosity depends on distance squared. An independent and precise distance measurement is important for many studies of NGC\,3783.

In the following sections, we describe our program to detect and characterize Cepheid variable stars to constrain the distance to NGC\,3783. Details of our observing plan, observations, and photometric analysis procedures are described in Section \ref{sec:obs}. Our criteria for selecting Cepheid candidates are outlined in Section \ref{sec:cand}. We fit the period$-$luminosity relationship to our Cepheid candidates in Section \ref{sec:pl} and provide calculations of distance moduli in \ref{sec:dm}. We discuss the potential effects of metallicity in Section \ref{sec:metal}. Previous distance measurements and implications of our distance measurement are discussed in Section \ref{sec:disc}. Lastly, Section \ref{sec:sum} provides a summary.

\section{Observations and Photometry} \label{sec:obs}

Multi-epoch observations of NGC\,3783 were obtained with the HST Wide Field Camera 3 (WFC3)\footnote{The specific observations analyzed can be accessed from MAST at the Space Telescope Science Institute via \dataset[doi: 10.17909/qe0p-a698]{https://doi.org/10.17909/qe0p-a698}.}, and they were planned as part of program GO-17165 (see also \citealt{markham_cepheid-based_2026}). Using the UVIS channel, we observed in three optical bandpasses: F350LP, F555W, and F814W. The field of view was $2.6' \times 2.6'$, and the pixel scale was $0.04 \arcsec$. We selected the UVIS-FIX aperture to hold the field center fixed, and we planned the visits with a single roll angle over the observation period. A four-point dither pattern was used to increase the sampling of the point spread function (PSF) and fill in the physical gap between the UVIS detectors.

Cepheid pulsations are stronger at bluer wavelengths, though extinction is also greater at these wavelengths \citep{freedman_hubble_1994}. The F555W filter retains a strong amplitude of variability while reducing effects of reddening. To obtain color information, we observed in the F814W filter as well. However, since NGC\,3783 is expected to lie at a distance of $\sim 30-40$ Mpc, we observed primarily in the long-pass optical filter F350LP. With a filter transmission that is broader than both F555W and F814W combined, the F350LP filter decreases the necessary observing time to detect and phase Cepheid candidates by a factor of $\sim 2.5$ \citep{hoffmann_optical_2016}.

Observations of NGC\,3783 spanned from April to August of 2023, for a total observation period of 127.76 days. Further details about our observing cadences are listed in Table \ref{tab:obslog}.

\begin{deluxetable*}{cccc}
    \tablewidth{0pt}
\tablecaption{Observations of NGC\,3783}
\tablehead{
\colhead{Visit} &
\colhead{UT} &
\colhead{Filter} &
\colhead{Exposure Time (s)}
}

\startdata
$1$ & 2023 Apr 15 & F350LP & 2260 \\
 & \nodata & F555W & 2240 \\
 & \nodata & F814W & 2240 \\
$2$ & 2023 Apr 27 & F350LP & 2260 \\
 & \nodata & F555W & 2240 \\
$3$ & 2023 May 6 & F350LP & 2260 \\
 & \nodata & F555W & 2240 \\
 & \nodata & F814W & 2240 \\
$4$ & 2023 May 13 & F555W & 2240 \\
$5$ & 2023 May 16 & F814W & 2240 \\
$6$ & 2023 May 25 & F350LP & 2260 \\
 & \nodata & F555W & 2240 \\
 & \nodata & F814W & 2240 \\
$7$ & 2023 May 31 & F814W & 2240 \\
$8$ & 2023 Jun 7 & F350LP & 2260 \\
 & \nodata & F814W & 2240 \\
$9$ & 2023 Jun 14 & F555W & 2240 \\
$10$ & 2023 Jun 20 & F814W & 2240 \\
$11$ & 2023 Jun 29 & F350LP & 2260 \\
 & \nodata & F555W & 2240 \\
$12$ & 2023 Jul 6 & F350LP & 2260 \\
 & \nodata & F555W & 2240 \\
 & \nodata & F814W & 2240 \\
$13$ & 2023 Aug 1 & F350LP & 2152 \\
$14$ & 2023 Aug 5 & F350LP & 2260 \\
$15$ & 2023 Aug 21 & F350LP & 2152 \\
\label{tab:obslog}
\enddata
\end{deluxetable*}

We planned 12 total visits to NGC\,3783 using the F350LP broadband filter, with eight additional observations in both the F555W and F814W filters distributed among the total visits. During the observation period, however, we encountered difficulties with guide star lock, resulting in several failed F350LP filter observations. Three of these failed observations were rescheduled for the end of the observing period, and we selected roll angles that most closely matched our previous roll angles. This resulted in ten viable observations of NGC\,3783 in the F350LP filter, and eight each observations in the F555W and F814W filters.

We downloaded the charge transfer efficiency-corrected images from the Mikulski Archive for Space Telescopes (MAST). These images were registered to a common frame and drizzled using the AstroDrizzle package \citep{fruchter_betadrizzle_2010}. The F350LP observations were stacked into one broadband ``deep frame", and individual visits were drizzled with respect to this frame. F555W and F814W observations were each stacked using the F350LP deep frame as reference, so point source coordinates were uniform between filters. Finally, the individual visits in F555W and F814W were all drizzled with respect to the F350LP deep frame.

We performed PSF photometry using the DAOPHOT package \citep{stetson_daophot_1987} on all point sources identified in our field. We followed a similar procedure to \cite{bentz_cepheid-based_2019} and \cite{markham_cepheid-based_2026}, with the addition of observations in F350LP as our new deep frame.

Using the F350LP deep frame image, 154,727 individual sources were detected in NGC\,3783. In creating star lists for our PSF models, we carefully selected bright sources without close neighbors and rejected extended sources, choosing 34 to create the PSF model for the F350LP deep frame image. This PSF model did not successfully fit sources in F350LP individual visits, so we created a separate star list using a sample of 36 bright sources in the first individual visit with the F350LP filter, with 16 sources common between the two star lists. We used this new sample to create PSF models for each individual visit in the F350LP filter, and we verified that the shapes of these models matched the shape of the deep frame PSF model. This procedure was repeated for the two remaining filters, although we were able to use the PSF star list created from the stacked images and did not need to create new lists for individual visits. In F555W, 35 sources were selected, and in F814W, 37 sources were selected. Using the sources in each star list, we created PSF models to carry out PSF photometry in all individual visits for each filter.

After measuring the fitted PSF magnitudes, we carried out aperture photometry on the bright sources that were used to create the PSF model. We then calculated the difference between the PSF magnitudes and the standard WFC3 10-pixel aperture magnitudes of the bright sources. For each individual visit in each filter, we accepted the median difference as the aperture correction, and we added these corrections to our PSF magnitudes for all point sources. These corrections were found to be approximately $-0.3$ mag in F350LP, $-0.2$ mag in F555W, and $-0.3$ mag in F814W. The aperture corrections were all negative and therefore yielded brighter corrected magnitudes.

Once the magnitudes were corrected, point sources were matched across all visits in the F350LP, F555W, and F814W filters. This produced 1541 unique point sources in NGC\,3783. We adopted VEGAMAG zero-points for a WFC3 10-pixel aperture in all three filters. These zero-points were 25.727 mag in F555W and 24.581 mag in F814W, given by \cite{riess_large_2019}. In F350LP, the zero-point was 26.693 mag, from the last set of tabulated zero-points for a 10-pixel aperture (\citealt{deustua_wfc3uvis_2017}; WFC3-ISR 2017-14). The typical $5\sigma$ depths in individual visits after corrections were $\sim26.8$ mag in F350LP, $\sim26.3$ mag in F555W, and $\sim25.1$ mag in F814W.

\section{Cepheid Candidate Selection} \label{sec:cand}

\begin{figure}[htb!]
    \centering
    \includegraphics[width=\columnwidth]{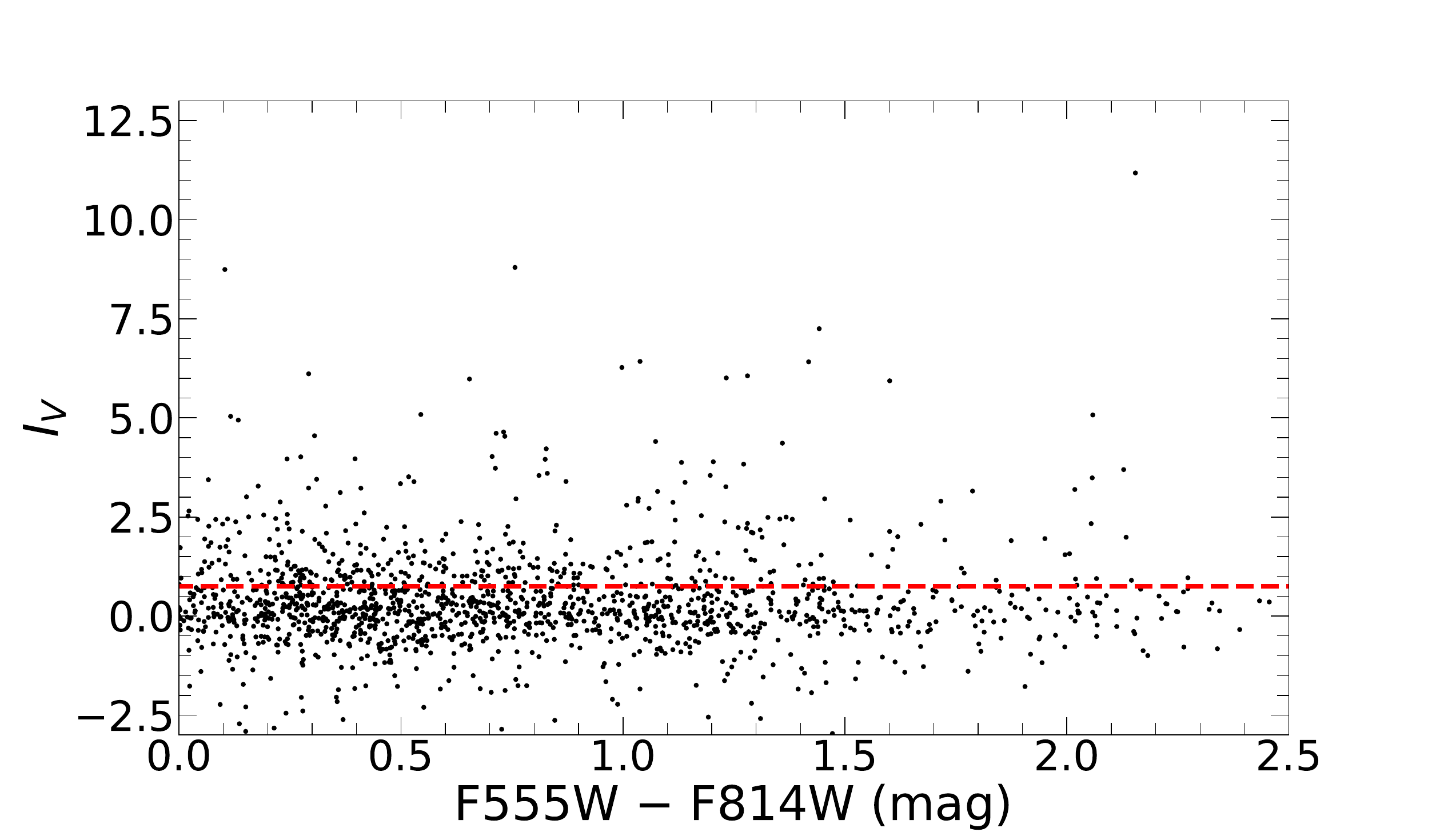}
    \caption{Variability index ($I_V$) shown against F555W$-$F814W color for all point sources in NGC\,3783. The red dashed line marks our selection for potentially variable sources at $I_V > 0.75$.}
    \label{fig:Ivar}
\end{figure}

To begin our search for Cepheid candidates in NGC\,3783, we first determined which sources exhibited evidence of variability. We calculated the \cite{welch_robust_1993} variability index for each matched point source, given by:
\begin{equation}
    I_{V} = \sqrt{\frac{1}{n(n-1)}} \sum^{n}_{i=1} \delta V_{i} \delta I_{i}
    \label{eq:var}
\end{equation} where $\delta V_{i}$ and $\delta I_{i}$ are the magnitude deviations in F555W and F814W, respectively, relative to the weighted mean magnitudes. Potentially variable sources should have greater variability indexes, while uncorrelated photometric errors should yield $I_V$ values around zero. We selected sources with $I_V > 0.75$ as our variable candidate sample, leaving us with 422 sources (see Figure \ref{fig:Ivar}). This limit has been relaxed compared to recent work by \cite{markham_cepheid-based_2026} to account for the greater distance and fainter stars in this galaxy. We also found that searching for variability with the F555W and F814W filters yielded more sources than any combination involving the F350LP filter, though $\sim140$ sources were common between all three combinations.

To determine which potential variable sources were Cepheid candidates, we fit the long-period Cepheid templates of \cite{yoachim_panoply_2009} to our light curve data. We fit all three filters simultaneously. Similar to prior work (e.g., \citealt{bentz_cepheid-based_2019}; \citealt{markham_cepheid-based_2026}), we created an array of 100 initial period guesses, spaced logarithmically from 10 days to 125 days. For each source, we ranked the fits based on each initial period guess by their $\chi^2$ values. Taking the fits with the six lowest $\chi^2$ values, we examined the best-fit periods. If the periods all agreed within 1.4 days, then the source was accepted as a Cepheid candidate. Our criterion of a 1.4-day range of agreement was selected to be similar to---without exceeding---the measurement uncertainties on the periods. This process was repeated for the entire variable candidate sample, and 115 sources passed this initial selection. For each of these sources, we recorded the best-fit light curve parameters: mean F350LP magnitude, mean F555W magnitude, mean F814W magnitude, period, phase, and all respective uncertainties. 

We then examined the $\chi^2$ values for the best fits of these sources and compared them with $\chi^2$ values for a straight-line fit in the F350LP filter, and we rejected sources that were better fit with a straight line than a Cepheid template. We also discarded candidates that did not have colors of $0.0 < \rm F555W - F814W < 2.0$ to remove extreme color outliers, compared to an expected color range of $0.5 < \rm F555W - F814W < 1.5$ for Cepheids on the instability strip \citep{shappee_new_2011}. A total of 20 sources passed these cuts, spanning a best-fit period range of $51.87 - 109.07$ days.

We visually inspected the remaining 20 sources. We searched for evidence of a distinctive ``sawtooth" profile in the F350LP light curves, since the F814W and F555W observations were shallower and more noisy. We determined that eight of these sources were likely Cepheid candidates.

At this point, we calculated corrections due to field crowding for our visually-determined Cepheid candidates. Following the procedures of \cite{bentz_cepheid-based_2019} and \cite{markham_cepheid-based_2026}, we used ADDSTAR to insert artificial stars within a $40 \times 40$ pixel box around each Cepheid candidate. This was done 100 times for each Cepheid candidate, and these artificial stars were randomly assigned F350LP, F555W, and F814W magnitudes within $\pm 0.5$ mag of the Cepheid candidates' measured magnitudes in a single visit. These artificial stars were used to determine the average impact of field crowding on the measured magnitudes. We performed photometry using DAOPHOT (as described in Section \ref{sec:obs}), and we determined the newly measured magnitudes of the Cepheid candidates and their real and artificial neighbors. We calculated the median difference between the assigned magnitudes of the artificial stars and their measured magnitudes for each Cepheid candidate to determine the effects of close neighbors on measured magnitudes. We then averaged these differences in each filter across our eight Cepheid candidates to determine magnitude corrections. In F350LP, the median difference ($m_{assigned} - m_{measured}$) was $0.016 \pm 0.054$ mag; in F555W, the median difference was $0.026 \pm 0.097$ mag; in F814W, the median difference was $0.027 \pm 0.087$ mag. The differences were all positive, so the measured magnitudes are brighter than the assigned magnitudes. Thus, field crowding biases magnitudes brighter. The magnitude corrections in each filter were added to the respective mean magnitudes of each Cepheid candidate.

We then examined the colors and amplitudes of the eight remaining Cepheid candidates. While we were initially lax with our allowed color range to find as many candidates as possible, we found one candidate that appeared to be a blue color outlier ($\rm F555W - F814W = 0.33 \pm 0.15$ mag). We tightened our range of allowed colors within uncertainties to $0.5 < \rm F555W - F814W < 2.0$, though we retained a relaxed upper limit to account for Galactic extinction along the line of sight to NGC\,3783. This cut removed a single source, bringing our sample down to seven Cepheid candidates. We also checked that the color uncertainty was not significant for any of these candidates, and all had color uncertainties $<0.2$ mag. Lastly, we examined the amplitudes of our seven candidates; we primarily examined the F350LP amplitudes. Following \cite{yuan_cepheid_2020}, we verified that the amplitude ranges of the F350LP light curves were within $0.4-1.7$ mag, and we also verified that all sources had F814W/F555W amplitude ratios $<1.0$. These checks did not remove any sources, and we retained our seven Cepheid candidates. 

\begin{figure*}[htb!]
    \centering
    \includegraphics[width=1.9\columnwidth]{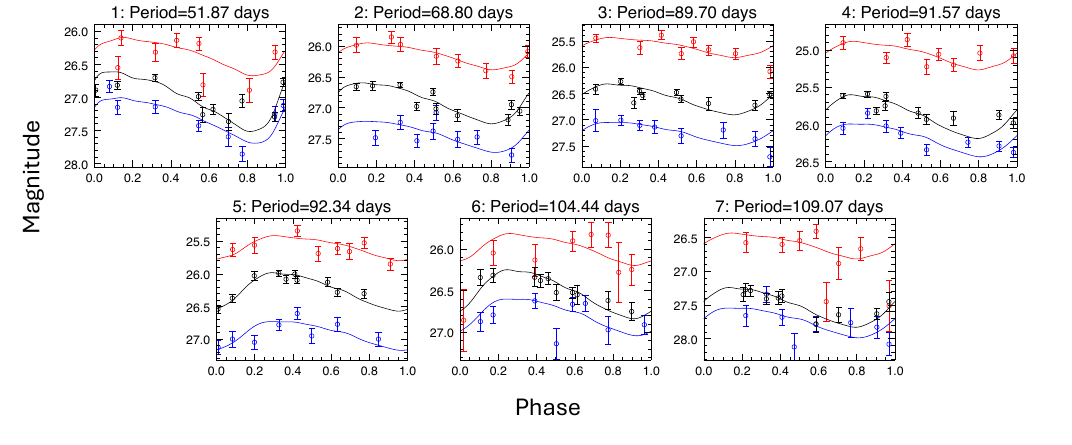}
    \caption{Best-fit light curves for our full sample of seven Cepheid candidates. Points mark our data in F350LP (black), F555W (blue), and F814W (red), while the solid curves represent the best fit returned by the Cepheid template fitting.}
    \label{fig:lcs}
\end{figure*}

Finally, we returned to visual inspection of our Cepheid candidates. Each candidate was given a grade: A, B, or C, based on the subjective quality of the fit. Grade A candidates show the strongest evidence for a sawtooth profile in the F350LP filter, and the scatter in the F555W and F814W filters is relatively low. Grade B candidates show strong evidence for a sawtooth profile, but they have comparatively greater scatter. Grade C candidates are the lowest quality sample---the sawtooth profile is not very distinct and/or the photometric errors and scatter are relatively large. We use this grading system to form two samples: our ``strong" Cepheid candidates solely comprised of grade As and Bs, and our full sample of all seven Cepheid candidates. The light curves of our full sample are displayed in Figure \ref{fig:lcs}, and the best-fit light-curve parameters are given in Table \ref{tab:ceph_fit}.

Given our average $5\sigma$ limiting magnitudes provided in Section \ref{sec:obs}, we expected to find Cepheid candidates with $P \gtrsim 90$ days. Most of our sample agreed with this limitation, though two sources in the strong sample were found with periods significantly below 90 days. Both of these sources (IDs $\#1$ and $\#2$) are located further from the center of NGC\,3783 and consequently have lower associated background levels, allowing them to be detected below the typical sensitivity achieved closer to the galaxy center.

\begin{deluxetable*}{rccccccccc}
    \tablecaption{Best-fit Light Curve Parameters for Cepheid Candidates in NGC\,3783}

\label{tab:ceph_fit}

\tablehead{
\colhead{ID} &
\colhead{RA} &
\colhead{Dec} &
\colhead{$m_{\rm F350LP}$} &
\colhead{$m_{\rm F555W}$} &
\colhead{$m_{\rm F814W}$} &
\colhead{$P$} &
\colhead{Phase} &
\colhead{$\chi^2 / \nu$} & 
\colhead{Grade} \\ 
\colhead{} &
\colhead{(hh:mm:ss)} &
\colhead{(dd:mm:ss)} &
\colhead{(mag)} &
\colhead{(mag)} &
\colhead{(mag)} &
\colhead{(days)} &
\colhead{(days)} &
\colhead{} &
\colhead{}
}

\startdata
1 & 11:38:57.680 & $-37:43:57.693$ & 27.02 ± 0.05 & 27.30 ± 0.07 & 26.34 ± 0.09 & 51.87 ± 0.81 & 23.66 ± 0.70 & 4.15 & B \\
2 & 11:39:06.653 & $-37:45:15.583$ & 26.89 ± 0.04 & 27.43 ± 0.06 & 26.14 ± 0.06 & 68.80 ± 2.05 & 46.57 ± 1.42 & 2.37 & A \\
3 & 11:39:03.305 & $-37:44:27.548$ & 26.55 ± 0.05 & 27.29 ± 0.09 & 25.66 ± 0.07 & 89.70 ± 5.49 & 0.00 ± 0.00 & 4.16 & A \\
4 & 11:39:00.621 & $-37:44:13.163$ & 25.88 ± 0.03 & 26.19 ± 0.05 & 25.06 ± 0.06 & 91.57 ± 5.19 & 83.41 ± 2.19 & 3.40 & C \\
5 & 11:39:01.049 & $-37:44:37.283$ & 26.26 ± 0.02 & 26.92 ± 0.04 & 25.59 ± 0.04 & 92.34 ± 3.65 & 88.75 ± 2.04 & 1.49 & A \\
6 & 11:39:00.463 & $-37:44:18.715$ & 26.53 ± 0.04 & 26.79 ± 0.05 & 25.99 ± 0.08 & 104.44 ± 9.20 & 75.77 ± 2.33 & 1.53 & C \\
7 & 11:39:06.500 & $-37:45:02.793$ & 27.52 ± 0.04 & 27.74 ± 0.08 & 26.61 ± 0.08 & 109.07 ± 12.29 & 85.25 ± 3.66 & 1.91 & C \\
\enddata
    \tablecomments{The ID's of each Cepheid candidate correspond to the ID's of each light curve in Figure \ref{fig:lcs}. The phases here are given in days with respect to the first visit. $\chi^2 / \nu$ is the reduced $\chi^2$ value for the F350LP Cepheid template fit. The grades A, B, and C correspond to visually-determined light curve quality assessments.}
\end{deluxetable*}

\section{Period$-$Luminosity Relationship} \label{sec:pl}

With our final list of Cepheid candidates, we fit the period$-$luminosity relationship to each sample based on the F555W, F814W, and Wesenheit magnitudes ($W_I$, \citealt{madore_period-luminosity_1982}). The Wesenheit magnitudes are described by:
\begin{equation}
    W_I = m_{\rm F814W} - R(m_{\rm F555W} - m_{\rm F814W})
    \label{eq:wes}
\end{equation} where $R = A_I / (A_V - A_I)$ ($R=1.3$ from \cite{cardelli_relationship_1989}). Wesenheit magnitudes are extinction-free and exhibit less scatter in their period$-$luminosity relationship when compared to relationships using F555W and F814W magnitudes \citep{madore_period-luminosity_1982}.

To fit the period$-$luminosity relationship, we used the calibrations by \cite{riess_large_2019} using LMC Cepheids for F555W, F814W, and $W_I$ magnitudes (also with $R=1.3$). These calibrations are based on a large sample and a precise independent distance constraint by \cite{pietrzynski_distance_2019}, and they removed the need to convert to \textit{V}- and \textit{I}-band magnitudes. The period$-$luminosity relationship is defined as:  
\begin{equation}
    m = \alpha \, \rm{log} \, \textit{P}\, + \beta
    \label{eq:pl}
\end{equation} where $m$ is the magnitude in each band, $\alpha$ represents the fixed slope from \cite{riess_large_2019} calibrations, $\beta$ is the intercept in magnitudes, and $P$ is the pulsation period in days. 

\begin{figure}[htb!]
    \centering
    \includegraphics[width=\columnwidth]{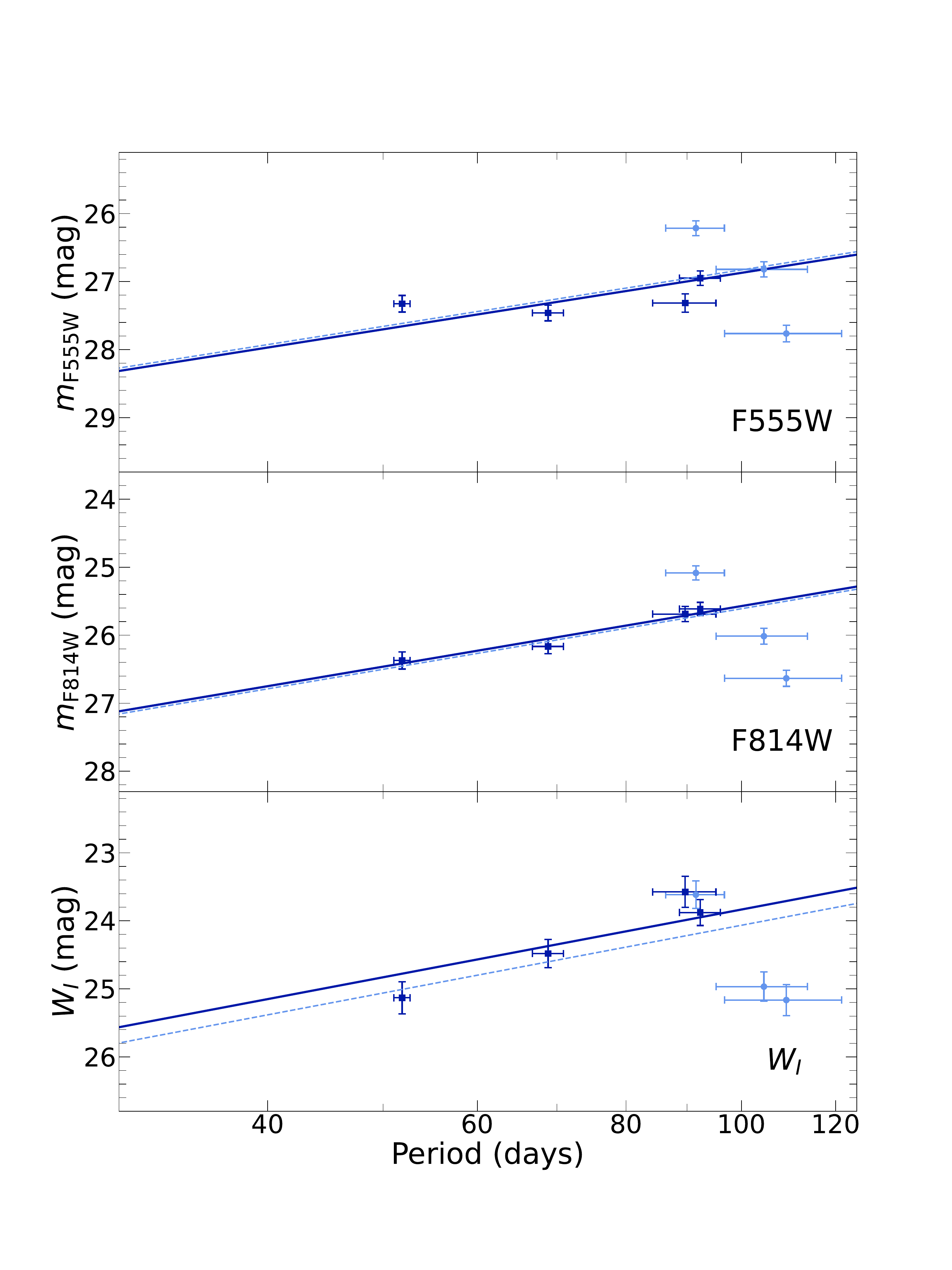}
    \caption{Period$-$luminosity relationship fits to our strong sample (dark blue squares) and our full sample (light blue circles) of Cepheid candidates. The three panels display the F555W (top), F814W (middle), and $W_I$ (bottom) magnitudes. The best fits to the strong sample are shown by the solid dark blue line, with the slopes being fixed to the calibrated values by \cite{riess_large_2019}. The light blue dotted lines display the best fits to the full sample for comparison. Corrections for Galactic extinction and metallicity have not been applied here.}
    \label{fig:PL}
\end{figure}

We fit both samples of Cepheid candidates using the SciPy Orthogonal Distance Regression routine \citep{virtanen_scipy_2020}. The slopes were held fixed to their calibrated values from \cite{riess_large_2019}, and the intercepts were allowed to vary. Accounting for both $x$- and $y$-dimensional uncertainties, this returned the best-fit parameters for the fitting function that was used---given by Equation \ref{eq:pl}. Additionally, we used the returned residual variance values to calculate $\sigma$ in each band. Period$-$luminosity relationships for both samples are shown in Figure \ref{fig:PL}.

Using the $\sigma$ values calculated in each band, we checked for the presence of $3\sigma$ outliers. There were no outliers in either sample, so we retain all the Cepheid candidates. We investigated the possibilities of blue blends as described by \cite{markham_cepheid-based_2026}, though we found none in our sample. Finally, due to the sparse nature of our sample, we declined potential cuts for completeness at shorter periods.

NGC\,3783 lies relatively close to the Galactic plane, and so our F555W and F814W magnitudes must be corrected for the effects of Galactic extinction. \cite{schlafly_measuring_2011} found Galactic extinction in the direction of NGC\,3783 to be $A_V = 0.346 \pm 0.040$ mag and $A_I = 0.186 \pm 0.040$ mag. These values were subtracted from the $\beta$ intercept values in the F555W and F814W filters, respectively. Our extinction-corrected intercepts are listed in Table \ref{tab:pl}.

\subsection{Distance Modulus} \label{sec:dm}

The distance modulus to the LMC is known within $1\%$ ($\mu_{\rm LMC} = 18.477 \pm 0.026$ mag) from observations of eclipsing binaries by \cite{pietrzynski_distance_2019}. We used this independent measurement to determine distance moduli for each of our Cepheid candidate samples in NGC\,3783. Using the $W_I$ magnitudes of our strong sample, we find $\mu = 32.996 \pm 0.152$ mag, and with the full sample, we find $\mu = 33.227 \pm 0.255$ mag. These correspond to distances of $D = 39.73 \pm 2.78$ Mpc and $D = 44.20 \pm 5.20$ Mpc, respectively. Full lists of our intercept values and distance moduli in each filter are given in Table \ref{tab:pl}. It is worth noting that the $W_I$ intercept value for the full sample is consistent within $1\sigma$ of the intercept found for the strong sample, so our choice of candidates does not strongly affect the distance measured for NGC\,3783.

\begin{figure}[htb!]
    \centering
    \includegraphics[width=\columnwidth]{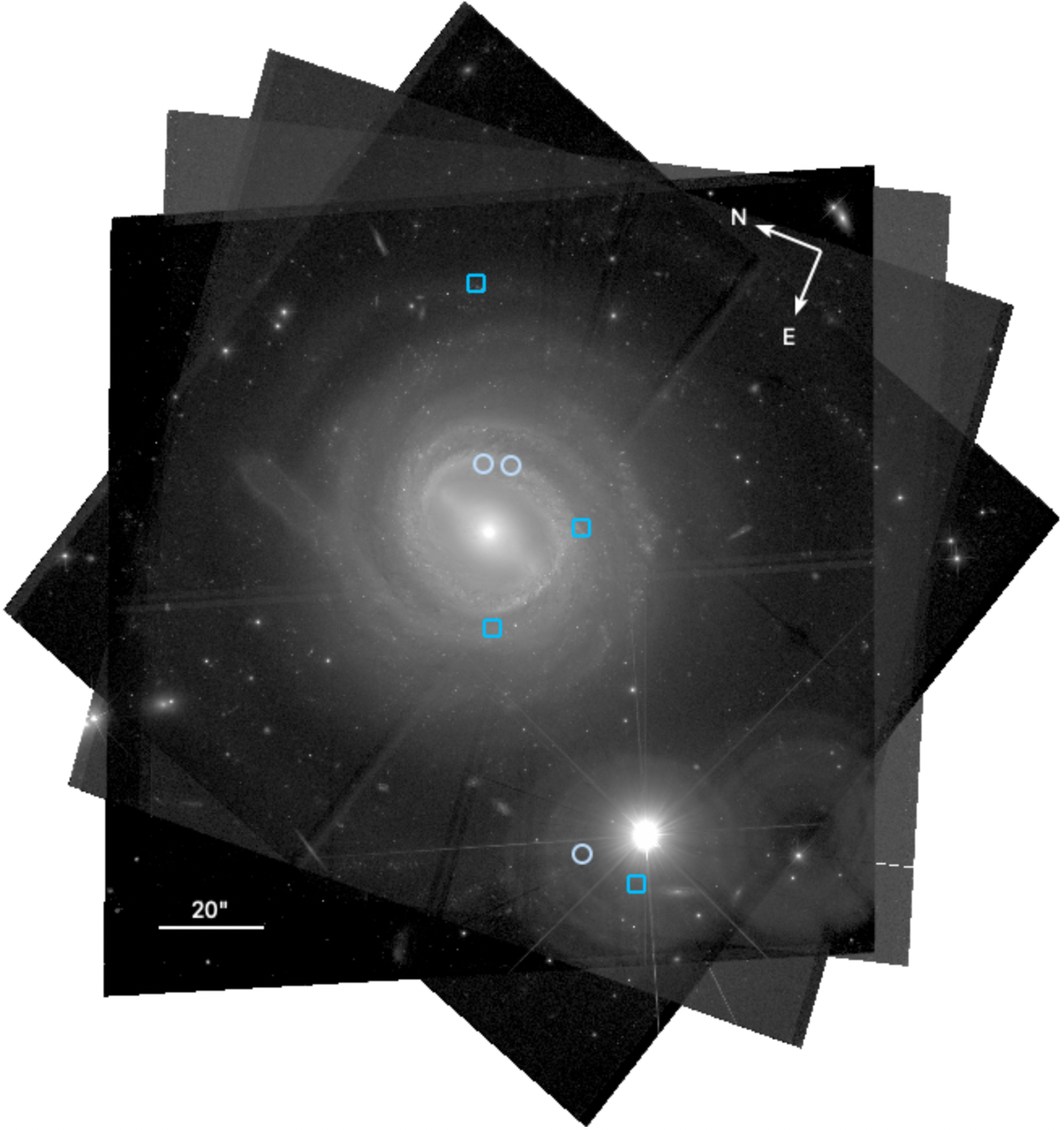}
    \caption{F350LP deep image of NGC\,3783 with the positions of our strong sample of four Cepheid candidates marked by blue boxes. The remaining three candidates in our full sample are marked by lighter blue circles. A compass is provided in the upper right corner, and a $20\arcsec$ scale bar is provided in the lower left.}
    \label{fig:positions}
\end{figure}

\begin{figure}[htb!]
    \centering
    \includegraphics[width=\columnwidth]{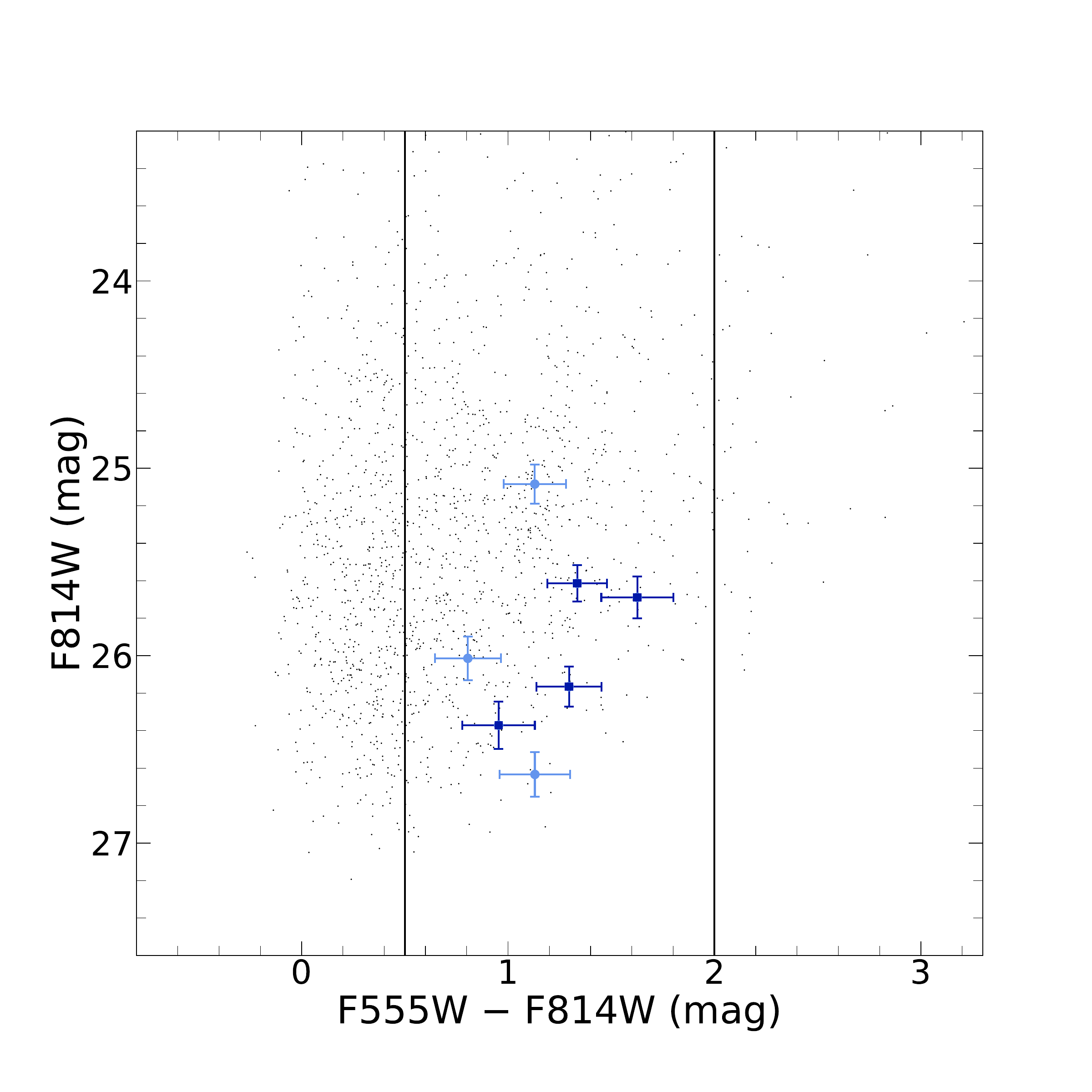}
    \caption{Color$-$magnitude diagram of all point sources detected in NGC\,3783 (shown in black). Our strong sample of Cepheid candidates are marked with dark blue squares, and additional light blue circles mark the remaining candidates in the full sample. Black vertical lines delimit our range of acceptable F555W$-$F814W colors.}
    \label{fig:cmd}
\end{figure}

\subsection{Metallicity Dependence} \label{sec:metal}

The strength of the effect of metallicity on the period$-$luminosity relationship is an ongoing area of study. The correlation factor between the period$-$luminosity relationship zero-point and metallicity ($\gamma$) has been calibrated many times using data from a variety of sources (see \citealt{markham_cepheid-based_2026} for a summary). Currently, most studies suggest that either metal-rich Cepheids are brighter and there is a negative correlation with metallicity ($\gamma < 0 \rm \ mag \ dex^{-1}$), or that there is no significant correlation ($\gamma \approx 0 \rm \ mag \ dex^{-1}$). Here, we investigate the potential effects of metallicity on the period$-$luminosity relationship for NGC\,3783. 

\cite{breuval_small_2024} present the following correction to the distance modulus relative to the LMC:
\begin{equation}
    \mu_{\rm rel \: LMC} = (\beta_{\rm NGC\,3783} - \beta_{\rm LMC}) + \Delta m
    \label{eq:metal_modulus}
\end{equation} where $\beta_{\rm NGC\,3783}$ is our best-fit intercept of the period$-$luminosity relationship using $W_I$ magnitudes, and $\beta_{\rm LMC}$ is the intercept for the LMC period$-$luminosity fit calibrated by \cite{riess_large_2019}. Here, $\Delta m$ represents a correction for the difference in average metallicity between LMC Cepheids and our Cepheid candidates in NGC\,3783:
\begin{equation}
    \Delta m = -\gamma ([\rm O/H]_{NGC\,3783} - [O/H]_{LMC})
    \label{eq:metal_calibration}
\end{equation} where $\gamma$ is a metallicity correlation factor and is calibrated as $\gamma = -0.264 \pm 0.058  \rm \ mag \ dex^{-1}$ \citep{breuval_small_2024}. We note that similar values for $\gamma$ have been found using both [O/H] and [Fe/H] metallicities (i.e., \citealt{riess_comprehensive_2022,breuval_small_2024}). The average metallicities of Cepheid candidates in NGC\,3783 and in the LMC are given by $[\rm O/H]_{NGC\,3783}$ and $[\rm O/H]_{LMC}$, respectively. Regarding the LMC metallicity, we used the value determined by \cite{romaniello_iron_2022} of $[\rm O/H]_{\rm LMC}=-0.32 \pm 0.01$, relative to a solar abundance of $12 + \rm log([O/H]) = 8.69$ \citep{asplund_chemical_2009}.

\begin{figure}[htb!]
    \centering
    \includegraphics[width=\columnwidth]{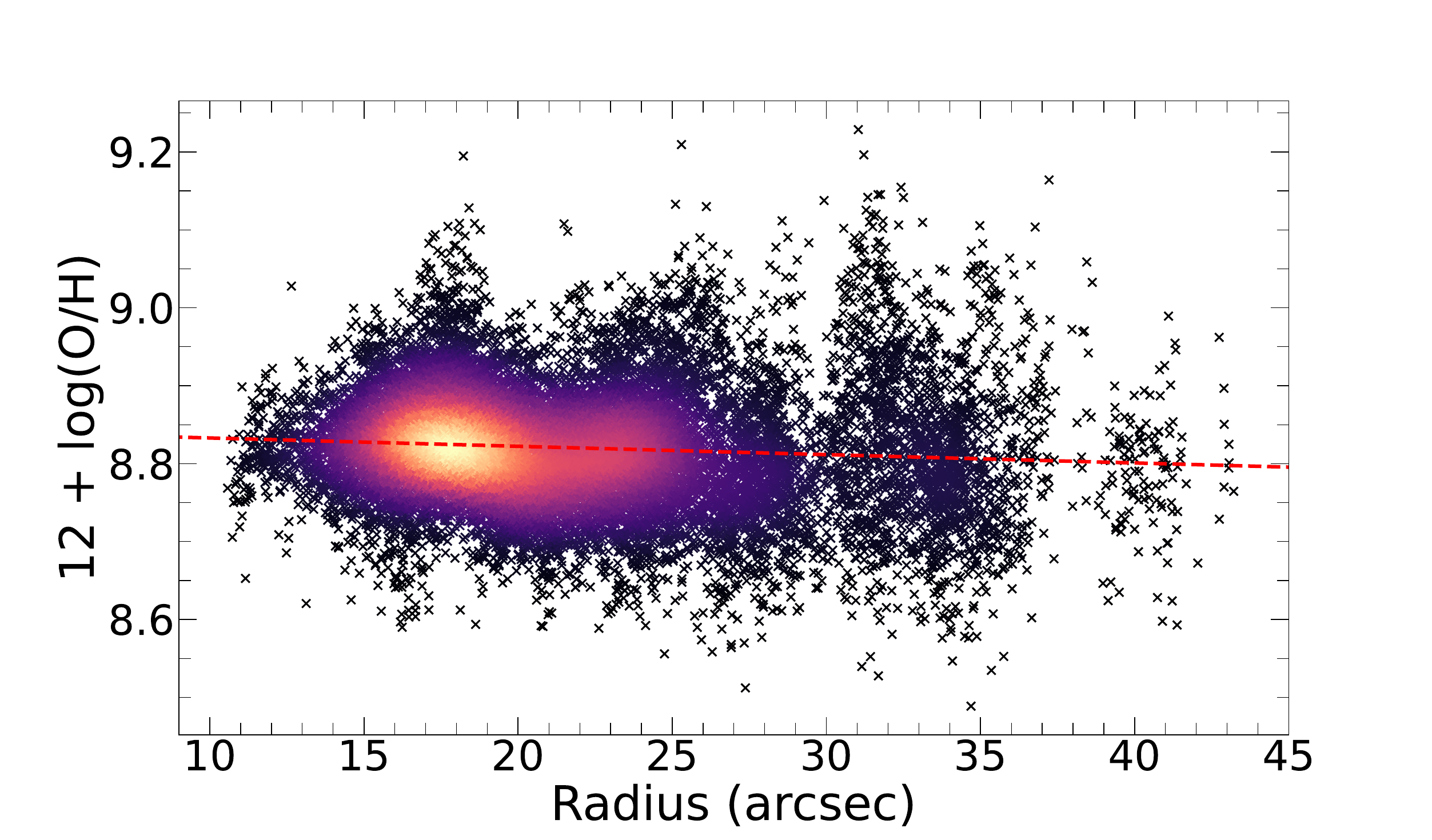}
    \caption{Radial metallicity gradient for NGC\,3783 produced from the 2D gas metallicity map by \cite{erroz-ferrer_muse_2019}. Individual metallicity measurements from \ion{H}{2} regions are marked by x's, and the density of points is illustrated by a color gradient where higher density regions are shaded yellow. The red line denotes the average metallicity as a function of radial distance.}
    \label{fig:met_grad}
\end{figure}

The MUSE Atlas of Discs (MAD) survey provides spatially-resolved ionized gas kinematics in nearby star-forming galaxies using VLT/MUSE data. \cite{erroz-ferrer_muse_2019} produced a 2D gas metallicity map of star forming regions in NGC\,3783. They determine the $12+ \rm log([O/H])$ metallicity using H$\alpha$, [\ion{N}{2}], and [\ion{S}{2}] emission lines in \ion{H}{2} regions and the calibration of \cite{dopita_chemical_2016}, which these authors determined to be the most robust when compared to two additional calibrations. We take these metallicity values and determine their deprojected spatial positions relative to the center of the galaxy (assuming $i = 25 \degree$, derived from \ion{H}{1} gas kinematics; \citealt{garcia-barreto_hi_1999}). Using these values, we calculate a radial metallicity gradient, shown in Figure \ref{fig:met_grad}. We find a central metallicity of $12+ \rm log([O/H]) = 8.86$ and a radial gradient of $-1.1 \pm 0.1 \times 10^{-3} \rm \ dex \ arcsec^{-1}$. It is worth noting that the metallicity gradient we adopt here is determined from the metallicity of the ISM, which may be enriched relative to individual Cepheids via stellar winds or supernovae. However, it is not feasible to measure the metallicities of individual stars at the distance of NGC\,3783, so in the absence of individual stellar metallicities, this gradient is our best approximation.

To determine the average metallicity of our Cepheid candidates in NGC\,3783, we first calculated the deprojected angular separations of each candidate in our full sample relative to the center of the galaxy (again using $i=25 \degree$). Using our calculated radial metallicity gradient of $-1.1 \pm 0.1 \times 10^{-3} \rm \ dex \ arcsec^{-1}$, we found the average [O/H] metallicity at the location of each Cepheid candidate. Relative to solar, these values spanned $0.07-0.15$ dex and had a median of $0.14 \pm 0.03$ dex. Using Equation \ref{eq:metal_calibration}, this yielded a median metallicity correction of $0.12 \pm 0.03$ mag. We noted that three of our Cepheid candidates had metallicity determinations available at their corresponding locations within the 2D gas metallicity map by \cite{erroz-ferrer_muse_2019}. All three Cepheid candidates had slightly lower metallicity values from the map than were predicted by the radial gradient, with a median difference of $0.056$ dex. However, we do not have corresponding metallicity measurements at the locations of the other four Cepheid candidates, and the difference between the measurements at the Cepheid locations vs. the value predicted by the gradient is on the order of the uncertainties. 

The correction factor of $\Delta m = 0.12 \pm 0.03$ mag was added to the relative distance modulus determined using the $W_I$ magnitudes, as described by Equation \ref{eq:metal_modulus}. The same metallicity correction factor was applied to both our strong sample and our full sample of Cepheid candidates. Final metallicity-corrected distance moduli can be found in Table \ref{tab:pl}.

Correcting for the potential effects of metallicity, we measure distances to NGC\,3783 of $D = 42.03 \pm 2.98$ Mpc based on our strong sample and $D = 46.75 \pm 5.53$ Mpc based on our full sample. The distance measurement that we accept as our final value here is $D = 42.03 \pm 2.98$ Mpc, using the result of the metallicity-corrected $W_I$ magnitudes for the strong sample.

\begin{deluxetable*}{lccccc}
    \tablewidth{0pt}
\tablecaption{Period$-$Luminosity Relationships}
\tablehead{
\colhead{Band} &
\colhead{$\alpha$} &
\colhead{$\beta$} &
\colhead{$\sigma$} &
\colhead{$\mu_{\rm rel \: LMC}$ (mag)} &
\colhead{$\mu$ (mag)} 
}

\startdata
\multicolumn{6}{c}{Strong Sample} \\
\hline
F555W & $-2.76$ & $32.391 \pm 0.131$ & $1.194$ & $14.407 \pm 0.137$ & $32.883 \pm 0.139$ \\
F814W & $-2.96$ & $31.491 \pm 0.042$ & $0.401$ & $14.451 \pm 0.058$ & $32.928 \pm 0.063$ \\
$W_{I}$ & $-3.31$ & $30.454 \pm 0.149$ & $0.779$ & $14.518 \pm 0.149$ & $32.996 \pm 0.152$ \\
Metallicity-corrected $W_{I}$ & \nodata & \nodata & \nodata & \nodata & $\mathbf{33.118 \pm 0.154}$ \\
\hline
\multicolumn{6}{c}{Full Sample} \\
\hline
F555W & $-2.76$ & $32.348 \pm 0.183$ & $1.464$ & $14.364 \pm 0.187$ & $32.841 \pm 0.189$ \\
F814W & $-2.96$ & $31.530 \pm 0.172$ & $1.412$ & $14.490 \pm 0.177$ & $32.967 \pm 0.179$ \\
$W_{I}$ & $-3.31$ & $30.685 \pm 0.254$ & $1.189$ & $14.750 \pm 0.254$ & $33.227 \pm 0.255$ \\
Metallicity-corrected $W_{I}$ & \nodata & \nodata & \nodata & \nodata & $33.349 \pm 0.257$ \\
\enddata
    \tablecomments{Slopes ($\alpha$) were held fixed to the values determined from LMC Cepheids by \cite{riess_large_2019}. LMC-relative distance moduli ($\mu_{\rm rel \: LMC}$) here have been corrected for Galactic extinction as discussed in Section \ref{sec:pl}, but they have not been corrected for potential effects of metallicity. Distance moduli ($\mu$) were determined using an independent distance constraint to the LMC by \cite{pietrzynski_distance_2019}. Our final adopted distance modulus for the strong sample, after correction for metallicity effects, is provided in bold text. Separate calculations are provided for both our strong sample and our full sample.}
    \label{tab:pl}
\end{deluxetable*}

\section{Discussion} \label{sec:disc}

The results of this work place NGC\,3783 among the most distant galaxies measured using Cepheids. \cite{riess_comprehensive_2022} presented Cepheid distances to 37 nearby galaxies using HST observations gathered over 20 years and with more than 1000 orbits. NGC\,3783 has a distance modulus that is greater than all but four of the galaxies in their sample. Our Cepheid-based distance to NGC\,3783 is representative of the impressive capabilities of HST, as well as the limit of its reach.

There are only a few existing independent distance measurements for NGC\,3783, and most are based on the Tully-Fisher relationship. This method is not accurate for face-on galaxies, because it relies on correcting the line-of-sight velocity for the effect of inclination. \cite{tully_catalog_1988} determined a distance of $38.5 \pm 14.2$ Mpc, and $D=38.5$ Mpc is frequently adopted as the distance to NGC\,3783 in a wide range of studies on AGN structure (e.g., \citealt{gravity_collaboration_central_2021}) and feeding and feedback processes (e.g., \citealt{davies_insights_2015,de_marco_incoherent_2020,shablovinskaia_alma_2025}). More recently, \cite{robinson_tully-fisher_2021} used the Tully$-$Fisher method to derive a distance of $49.8 \pm 10.0$ Mpc from new observations of the \ion{H}{1} emission line and careful consideration of the AGN contribution to the host-galaxy magnitude. However, the nearly face-on inclination of NGC\,3783 ($i \approx 25 \degree$; \citealt{garcia-barreto_hi_1999}) leads to a large uncertainty in inclination correction for the \ion{H}{1} line width. Our Cepheid-based distance measurement of $D = 42.03 \pm 2.98$ Mpc agrees within the large $1\sigma$ uncertainties of both of these measurements, though we provide a significantly tighter constraint.

In addition to Tully$-$Fisher measurements, there are geometric distances available for NGC\,3783. \cite{pedreros_metric_1981} described an observational trend that, among galaxies with inner rings, more luminous galaxies have larger rings and that ring size decreases for later-type galaxies. Using the apparent inner ring diameter, they derive a distance of $40$ Mpc, assuming that $H_0 = 100 \rm \ km \ s^{-1} \ Mpc^{-1}$. If we instead adopt $H_0 = 73 \rm \ km \ s^{-1} \ Mpc^{-1}$ \citep{riess_comprehensive_2022}, this yields a distance of $55$ Mpc. This value lies at the upper end of reported distances for NGC\,3783, and it is a factor of $\sim 1.3$ times larger than our reported distance of $D = 42.03 \pm 2.98$ Mpc. Recently, \cite{gravity_collaboration_geometric_2021} produced a geometric distance measurement for NGC\,3783 using interferometric observations and existing RM data. The GRAVITY spectrometer on the Very Large Telescope Interferometer (VLTI) combines light from all Unit Telescopes and can resolve structures on a sub-milliarcsecond scale \citep{gravity_collaboration_first_2017}. \cite{gravity_collaboration_central_2021} presented a detailed view of the inner structure of NGC\,3783 and characterized its BLR as a rotating, thick disc with a mean radius of $R_{\rm BLR} = 71 \ \mu \rm as$ using the Br$\gamma$ emission line. Combining the physical size of the BLR measured from RM with the angular size determined from interferometry, \cite{gravity_collaboration_geometric_2021} determined a geometric distance to NGC\,3783 of $D = 39.9^{+14.5}_{-11.9}$ Mpc. The uncertainties on this measurement are quite large, and our Cepheid-based distance reduces the relative uncertainties from $\sim 36\%$ to $\sim 7\%$.

Furthermore, NGC\,3783 has been described in the literature as being part of several different galaxy groups with varying average distances. The grouping of \cite{kollatschny_group_1989} placed NGC\,3783 in a group of three galaxies (along with NGC\,3749 and NGC\,3742) with an average distance of $36.9$ Mpc, determined from the average redshift of the galaxies and using $H_0 = 75 \rm \ km \ s^{-1} \ Mpc^{-1}$. \cite{crook_groups_2007} placed it in a four-member group with distances ranging from $35.96-37.64$ Mpc and with a mean distance of $36.76$ Mpc. Distances to group members in this catalog were derived from redshifts after correcting for distortions to local velocity fields. More recently, \cite{kourkchi_galaxy_2017} placed it within a group of nine where two group members have non-redshift distances, $39.81$ Mpc and $43.85$ Mpc, averaging to $42.08 \pm 2.10$ Mpc. Finally, \cite{kilborn_gaseous_2006} place NGC\,3783 in a group of 47 members based on a spatial cut and a velocity selection of $v = 2500-3500 \rm \ km \ s^{-1}$. Within the group of 47 galaxies, 12 had distinct \ion{H}{1} detections. For a group distance, however, \cite{kilborn_gaseous_2006} adopted $D = 36$ Mpc from \cite{osmond_gems_2004}, which used a group of three galaxies determined by \cite{giuricin_nearby_2000}. Our Cepheid-based distance of $D = 42.03 \pm 2.98$ Mpc, while in close agreement with the group distance from \cite{kourkchi_galaxy_2017}, is $\sim 1.2$ times greater than a group distance of $36$ Mpc. The distance measurement we provide here will enable future studies of NGC\,3783 to more accurately understand its context among neighbor galaxies.

Using our Cepheid-based distance measurement, we determined the peculiar velocities affecting NGC\,3783. Given an \ion{H}{1} recessional velocity of $2917 \rm \ km \ s^{-1}$ \citep{theureau_kinematics_1998} and $H_0 = 73 \rm \ km \ s^{-1} \ Mpc^{-1}$ \citep{riess_comprehensive_2022}, we calculated a redshift distance of $D = 40.0$ Mpc. Assuming the distance given by our strong sample of $D = 42.03 \pm 2.98$ Mpc, we determined approximate peculiar velocities of $-100$ to $-200 \rm \ km \ s^{-1}$. \cite{gravity_collaboration_geometric_2021}, who determined a geometric distance to NGC\,3783 of $39.9^{+14.5}_{-11.9}$ Mpc, also investigated the peculiar velocities of NGC\,3783. They calculated a velocity of $2628 \rm \ km \ s^{-1}$ for NGC\,3783 in the reference frame of the Local Sheet instead of using a heliocentric velocity. Using the peculiar velocity calculation from \cite{davis_deriving_2014} and assuming $H_0 = 70 \rm \ km \ s^{-1} \ Mpc^{-1}$, they also found peculiar velocities around $-100$ to $-200 \rm \ km \ s^{-1}$. Peculiar velocities on the order of a few hundred kilometers per second are not unusual for nearby galaxies, especially those within groups or clusters \citep{tully_cosmicflows-2_2013}. Given the various groups that NGC\,3783 has been considered a member of, it is possible that some of its neighbors are affecting its \ion{H}{1} recessional velocity, and this underscores that redshift alone is not a sufficient distance indicator for nearby AGN.

While NGC\,3783 has a black hole mass measurement that is distance-independent, other physical parameters are affected by our new distance measurement. \cite{davies_insights_2015} determined a bolometric luminosity of $\rm log \ \it{L_{\rm AGN}} \rm \approx 44.5 \ erg \ s^{-1}$ assuming $D=38.5$ Mpc, and our reported distance of $D = 42.03 \pm 2.98$ Mpc implies that the AGN luminosity is $\sim 1.2$ times greater. \cite{mehdipour_chasing_2017} and \cite{mao_photoionized_2019} each produced an SED for NGC\,3783, and they both adopted its redshift ($z = 0.00973$; \citealt{theureau_kinematics_1998}) instead of an independent distance measurement. In this case, our Cepheid-based distance yields a luminosity that is $\sim 1.1$ times greater than predicted by the redshift. The SED from \cite{mehdipour_chasing_2017} has been used in recent studies on the X-ray outflows in NGC\,3783 (e.g., \citealt{mehdipour_delving_2025,zhao_delving_2026}), and it was specifically used to model obscuring winds that have exhibited change over time. Future modeling of these outflows will benefit from an updated SED that adopts our Cepheid-based distance measurement.

\section{Summary} \label{sec:sum}

We have presented a multi-epoch HST program to detect and characterize Cepheids in the nearby active galaxy NGC\,3783. We identified Cepheid candidates and fit the period$-$luminosity relationship to our best candidates as well as to all the candidates and determined the resultant distance moduli. Correcting for metallicity, our strong sample of Cepheid candidates yielded $\mu = 33.118 \pm 0.154$ mag and a distance of $D = 42.03 \pm 2.98$ Mpc. Thus, NGC\,3783 is one of the farthest galaxies to have a distance measurement based on Cepheids.

Our Cepheid-based distance formally agrees with prior distance measurements based on the Tully$-$Fisher relationship ($D = 38.5 \pm 14.2$ Mpc; $D = 49.8 \pm 10.0$ Mpc) and the geometric distance using interferometry ($D = 39.9^{+14.5}_{-11.9}$ Mpc), however, we provide a much tighter constraint. The uncertainties attributed to previous measurements are greater by factors of $\sim 3-5$ than what we present here. The distance commonly used to determine AGN bolometric luminosity ($D = 38.5$ Mpc) is $\sim 1.1$ times smaller than our distance, which corresponds to a factor of $\sim 1.2$ increase in luminosity by adopting our new distance. We found peculiar velocities around $-100$ to $-200 \rm \ km \ s^{-1}$ affecting NGC\,3783 and contributing to its observed redshift. Future studies of NGC\,3783 will benefit from the accuracy and substantially improved precision of our new distance measurement.

\begin{acknowledgments}
We thank D.\ Michael Crenshaw, Monica Valluri, and Eugene Vasiliev for their contributions to the proposal. M.\ Markham and M.C.\ Bentz gratefully acknowledge support through program HST GO-17165, which was provided by NASA through a grant from the Space Telescope Science Institute, which is operated by the Association of Universities for Research in Astronomy, Inc., under NASA contract NAS 5–26555. M.\ Vestergaard gratefully acknowledges financial support from the Independent Research Fund Denmark via grant number DFF 3103-00146 and from the Carlsberg Foundation (grant CF23-0417).
\end{acknowledgments}

\begin{contribution}
M.\ Markham drizzled the observations, carried out the analysis, and wrote the manuscript. M.C.\ Bentz led the proposal and Phase 2 planning and oversaw the project. L.\ Ferrarese, C.A.\ Onken, and M.\ Vestergaard provided scientific expertise. All authors contributed to the presentation of the results.
\end{contribution}

\facility{HST (WFC3)}

\software{AstroDrizzle \citep{fruchter_betadrizzle_2010}, DAOPHOT \citep{stetson_daophot_1987}}

\bibliography{cepheids_ref_2}{}
\bibliographystyle{aasjournalv7}

\end{document}